\pdfoutput=1
\documentclass[jhep,preprint,floats,12pt,floatfix,superscriptaddress,preprintnumbers,showpacs,eqsecnum,
nofootinbib]{revtex4-1}
\usepackage[utf8]{inputenc}
\usepackage{latexsym,array,theorem,mathrsfs,bm,float}
\usepackage{amsfonts,amsmath,amssymb,latexsym,array,afterpage,
theorem,mathrsfs,bm,float,epsfig,color,
graphicx,tabularx,here,multirow,hyperref}
\usepackage{soul}

\newcommand{\nn}{\nonumber \\}

\newcommand{\Nold}{\tilde{N}}
\newcommand{\gamold}{\tilde{\gamma}}
\newcommand{\piold}{\tilde{\pi}}

\newcommand{\phiold}{\tilde{\phi}}

\begin{document}
\baselineskip=17pt

\begin{flushright}
YITP-21-24\qquad IPMU21-0021
\end{flushright}
\preprint{}
\title{Non-uniqueness of massless transverse-traceless graviton}
\author{Katsuki Aoki}
\email{katsuki.aoki@yukawa.kyoto-u.ac.jp}
\affiliation{Center for Gravitational Physics, Yukawa Institute for Theoretical Physics, Kyoto University, 606-8502, Kyoto, Japan}

\author{Francesco Di Filippo}
\email{francesco.difilippo@yukawa.kyoto-u.ac.jp}
\affiliation{Center for Gravitational Physics, Yukawa Institute for Theoretical Physics, Kyoto University, 606-8502, Kyoto, Japan}

\author{Shinji Mukohyama}
\email{shinji.mukohyama@yukawa.kyoto-u.ac.jp}
\affiliation{Center for Gravitational Physics, Yukawa Institute for Theoretical Physics, Kyoto University, 606-8502, Kyoto, Japan}
\affiliation{Kavli Institute for the Physics and Mathematics of the Universe (WPI), The University of Tokyo, Kashiwa, Chiba 277-8583, Japan}

\date{\today}

 \begin{abstract}
We study a theory of minimally modified gravity called cuscuton/VCDM that propagates only two gravitational degrees of freedom. Despite being apparently different from general relativity (GR), it is in principle possible that this theory might be obtained via a field redefinition starting from the GR action. This would make the vacuum theory equivalent to GR and the theory would differ from GR only in the presence of matter. In this paper, studying the dispersion relation of gravitational waves and the dynamics of the Bianchi-I universe, we prove that such a field redefinition does not exist and that the theory differs from GR already in vacuum.
 \end{abstract}

\maketitle

\section{Introduction}
General Relativity (GR) possesses several aspects that makes it a very simple and elegant theory. One such aspect is that it is a minimal theory of the gravitational interaction, meaning that it does not propagate any degrees of freedom other than the two degrees of freedom associated with the massless graviton.

It is interesting to study if there are other minimal theories of gravity without any extra degrees of freedom. On the one hand, GR is the unique theory of a transverse-traceless graviton having only two local degrees of freedom under some assumptions which will be detailed in Sec.~\ref{sec_unique}. On the other hand, Ref.~\cite{Lin:2017} introduced a class of minimal theories dubbed minimally modified gravity (MMG) which are only invariant under spatial diffeomorphism and time reparametrization and that propagates only two local degrees of freedom. Some of these theories are actually equivalent to GR \cite{Lin:2019,Carballo-Rubio:2018}. We call these theories type-I \cite{Aoki:2018}. Any type-I MMG theory can be obtained from GR by change of variables. In the absence of matter fields the type-I MMG theories are nothing but GR while the gravitational law is modified because of non-trivial matter coupling. Conversely, type-II MMG theories are characterized by the absence of an Einstein frame and they cannot be obtained from GR via a field redefinition. Therefore type-II MMG differ from GR also in vacuum.

In this paper we study a candidate type-II theory introduced in \cite{DeFelice:2020}. This theory is characterized by the presence of a function $V(\phi)$ of a non-dynamical auxiliary field $\phi$ that replaces the cosmological constant $\Lambda$. For this reason this theory, when combined with the CDM paradigm, is called VCDM~\cite{DeFelice:2020cpt}. In the present paper, although we shall not consider the CDM, we still call the candidate type-II theory VCDM. 
When $d^2V/d\phi^2=0$ the theory explicitly reduces to GR (see subsection~\ref{subsec:cuscuton-VCDM}). Although for a generic $V(\phi)$ the action of the theory is different from that of GR, the two theories would be physically equivalent in the absence of matter if there was a field redefinition that maps them to each other. The main purpose of this paper is to prove that such a field redefinition does not exist. 
To this end, in Sec.~\ref{sec:type-II} we show that gravitational waves propagating over a FLRW background in the candidate type-II theory have the same dispersion relation as in GR. We also show that the theory is equivalent to the cuscuton theory~\cite{Afshordi:2006ad}. We then study the dynamics of a Bianchi-I universe. In Sec.~\ref{sec:type-I} we start with GR and perform the most general local field redefinition that does not break spatial diffeomorphism invariance and that does not modify the dispersion relation of gravitational waves, thus obtaining the most general class of type-I MMG theories that have the same dispersion relation for gravitational waves. In principle, this class might contain our type-II candidate. We then study the Bianchi-I universe and we show that no theory in this class can reproduce the dynamics of the theory under consideration, thus showing that the latter is of type-II.

\section{On the uniqueness of general relativity}
\label{sec_unique}

Assuming the full diffeomorphism invariance, the second-order divergence-free field equations, and the absence of fields other than the metric, the Lovelock theorem assures that, in four-dimension, the only possible equations of motion are the Einstein field equations \cite{Lovelockth}. Given that higher-order field equations are generically associated with extra degrees of freedom, the Lovelock theorem can be considered as the first result on the uniqueness of GR as a theory describing only two degrees of freedom that correspond to the transverse-traceless graviton. However, this conclusion can be easily avoided by relaxing the assumption of the full diffeomorphism invariance. 

Another well known uniqueness result can be obtained following the analysis of \cite{Gupta,Deser1969}. The idea behind this approach is to recursively reconstruct the GR action starting from the linear action describing a massless graviton in flat spacetime. This result is not as general as it might seems. In fact, we are forced to make arbitrary choices at each step of the reconstruction procedure in order to obtain the GR action (see \textit{e.g} \cite{Padmanabhan2004,Carballo-Rubio:2018b}). These arbitrary choices are probably related to the freedom of performing an arbitrary field redefinition changing the action, but without changing the physical on-shell quantities. 

For these reasons, modern on-shell methods are better suited to discuss the constructability of the theory as they focus on the scattering matrix of the theory which is an on-shell quantity and it is invariant under field redefinition. 
With this formalism, a series of uniqueness results have been obtained \cite{Britto2005,Benincasa2007,ArkaniHamed2008,Carballo-Rubio:2018b,Pajer:2020}. All these works have a different set of assumptions leading to slightly different results. However, all of them assume locality and unitarity of the scattering matrix.
In particular, an essential property that is assumed in all the analyses above is that poles of the amplitudes correspond to internal particles going on-shell. This property is assured by the locality condition \cite{Benincasa:2013,Conde:2014}.

Therefore, while there are plenty of uniqueness results in the literature, the MMG theories that we are interested in in this paper avoid all of them since the locality condition does not hold here. 
Indeed, the study of spherically symmetric solutions uncovered the existence of a mode that follows an elliptic (instead of hyperbolic) equation~\cite{DeFelice:2020onz}. Such a mode is called a generalized instantaneous mode or a shadowy mode~\cite{DeFelice:2018ewo} and integrating it out results in a kind of non-local interaction. 
The Lovelock theorem cannot be applied as the theory is not invariant under the full diffeomorphism group, whereas the analyses based on the scattering amplitude matrix do not apply because the auxiliary field introduces a non-local interaction. 
In the absence of an uniqueness theorem that can be applied to the situation of interest, we will study a specific example of MMG theories to either find a canonical transformation to map it to GR, showing that it is a type-I theory, or to prove that such a field redefinition does not exist, showing that we are dealing with a type-II theory.

\section{Candidate type-II MMG}\label{sec:type-II}
\subsection{Cuscuton and VCDM theories}
\label{subsec:cuscuton-VCDM}

A candidate type-II MMG theory is the cuscuton theory~\cite{Afshordi:2006ad} which is a special class of k-essence theories. The Lagrangian is
\begin{align} \label{cuscuton_lag}
\mathcal{L}_{\rm cus}=\frac{1}{2}\sqrt{-g}\left[ R^{(4)}+ \sqrt{|(\partial \phi_{\rm cus})^2|}-2U(\phi_{\rm cus}) \right]
,
\end{align}
where $R^{(4)}$ is the four-dimensional Ricci scalar, $\phi_{\rm cus}$ is a cuscuton field and we have chosen the unit with $M_{\rm Pl}=1$. Although the theory apparently involves a scalar degree of freedom on top of the tensorial degrees of freedom, the Hamiltonian analysis proves that there are only two tensorial degrees of freedom around timelike configurations of $\partial_{\mu} \phi_{\rm cus}$ (see also~\cite{Iyonaga:2018vnu} for an extended version of the cuscuton theory). The cuscuton theory can be regarded as an infinite speed of sound limit of a k-essence theory which may evade the previous uniqueness results because a field with the infinite speed of sound introduces a non-local interaction. We also note that the cuscuton theory is a special class of the k-essence which possesses an infinite set of symmetries~\cite{Pajer:2018egx,Grall:2019qof}.

Another candidate theory is the VCDM model~\cite{DeFelice:2020,DeFelice:2020cpt,DeFelice:2020prd} of which gravitational Lagrangian is given by
\begin{align}\label{eq:type-II_lag}
\mathcal{L}_{\rm VCDM}=\frac{1}{2}N\sqrt{\gamma}\left[ R+K_{ij}K^{ij}-K^2-2V(\phi)-2\frac{\lambda^i_{\rm gf}}{N}\partial_i \phi-\frac{3\lambda^2}{2}-2\lambda(K+\phi) \right]\,.
\end{align}
Here, $R$ is the spatial Ricci scalar, $K_{ij}$ is the extrinsic curvature, and $\phi$, $\lambda$ are auxiliary variables.
The model is intended to realize a time-dependent dark energy component due to the breaking of four-dimensional diffeomorphism at large scales. The ``gauge-fixing'' term $\lambda^i_{\rm gf} \partial_i \phi$ imposes the condition that $\phi$ is a function of time. The auxiliary variables 
$\phi$, $\lambda$ can be integrated out. The Lagrangian can be then written by the ADM variables only,
\begin{align}
\mathcal{L}_{\rm VCDM}=\frac{1}{2}N\sqrt{\gamma}\left[ R+K_{ij}K^{ij}-K^2-2\mathscr{V}(K) \right]
\label{L_VCDM}
\end{align}
in the constant mean curvature slice $K=K(t)$, where
\begin{align}
\mathscr{V}=V-\frac{1}{3}(K+\phi)^2
\end{align}
and $\phi$ is understood as a root of
\begin{align}
V_{\phi}-\frac{2}{3}(K+\phi)=0
\,.
\end{align}
Here, $V_{\phi}=dV/d\phi$. 
The Hamiltonian analysis shows that \eqref{eq:type-II_lag} only has two tensorial degrees of freedom~\cite{DeFelice:2020}.
Also, if $V_{\phi\phi}=0$ then $\mathscr{V}=c_0+c_1K$, where $c_0$ and $c_1$ are constant, and thus the Lagrangian \eqref{L_VCDM} reduces to the GR Lagrangian up to total derivative. Here we have used $\frac{\partial}{\partial t}\sqrt{\gamma}=\frac{1}{2}\sqrt{\gamma}\gamma^{ij}\dot{\gamma}_{ij}$. 

Although the theories \eqref{cuscuton_lag} and \eqref{eq:type-II_lag} were independently introduced in different contexts, they are actually equivalent to each other. In order to show their equivalence, let us consider the unitary gauge of the cuscuton theory in which the Lagrangian is
\begin{align}
\mathcal{L}_{\rm cus}=\frac{1}{2}N\sqrt{\gamma}\left[ R+K_{ij}K^{ij}-K^2 + \frac{\dot{\phi}_{\rm cus}}{N} -2 U \right]
,
\end{align}
in the ADM form. By taking integration by parts and adding boundary terms, the Lagrangian is transformed into
\begin{align}
\mathcal{L}_{\rm cus}=\frac{1}{2}N\sqrt{\gamma}\left[ R+K_{ij}K^{ij}-K^2 - \phi_{\rm cus} K-2 U \right]
,
\end{align}
where we have used $\frac{\partial}{\partial t}\sqrt{\gamma}=\frac{1}{2}\sqrt{\gamma}\gamma^{ij}\dot{\gamma}_{ij}$ and $\phi_{\rm cus}=\phi_{\rm cus}(t)$. The variable $\phi_{\rm cus}$ is manifestly an auxiliary variable and can be integrated out. Then, we find
\begin{align}
\mathcal{L}_{\rm cus}=\frac{1}{2}N\sqrt{\gamma}\left[ R+K_{ij}K^{ij}-K^2 -2 \mathscr{U}(K) \right]
,
\label{L_cus}
\end{align}
in the constant mean curvature slice $K=K(t)$, where
\begin{align}
\mathscr{U}=U+ \phi_{\rm cus}K
\end{align}
and $\phi_{\rm cus}$ is a root of
\begin{align}
\frac{dU}{d\phi_{\rm cus}}+\frac{1}{2}K=0
\,.
\end{align}
The unitary gauge Lagrangian of the cuscuton is exactly the same as the gravitational Lagrangian of the VCDM model. In the argument above, we have implicitly assumed that the auxiliary variable $\phi$ in the VCDM and $\phi_{\rm cus}$ in the cuscuton are solvable in terms of $K$.

The Lagrangian of the cuscuton/VCDM is clearly different from that of GR if $\mathscr{U}=\mathscr{V}\neq c_0 +c_1 K$. However, it is too early to conclude that the cuscuton/VCDM is inequivalent to GR in the light of the uniqueness results of GR. There could be a non-trivial map between the cuscuton/VCDM and GR, meaning that on-shell quantities, say scattering amplitudes of the graviton-graviton scattering could be the same. If this is indeed the case, the cuscuton/VCDM is a type-I MMG. Comparing scattering amplitudes is a definitely robust check, but the calculations may be performed only perturbatively. For instance, the model $\mathscr{U}=\mathscr{V} \propto K^n$ with a natural number $n$ should predict the same scattering amplitudes as GR up $(n-1)$-point amplitudes and the first deviation should appear from $n$-point amplitudes.

In the following, we will study the dispersion relation of gravitational waves and the dynamics of the Bianchi I universe. The shear of the spacetime can be interpreted as the zero momentum mode of gravitational waves, meaning that the dynamics of the Bianchi I universe is analogous to a ``scattering'' of gravitons including all the backreactions (needless to say, the notion of scattering does not make sense here). Our strategy is to compare the dynamics of the Bianchi I universe in the cuscuton/VCDM with that in the most general type-I MMG theory, the latter of which will be introduced in Sec.~\ref{sec:type-I}.

\subsection{Gravitational waves}

From Eq.~\eqref{L_VCDM} and \eqref{L_cus}, one sees that the Lagrangian consists of the combination $R+K_{ij}K^{ij}-\frac{1}{3}K^2$ and an arbitrary function of $K$. The trace of the extrinsic curvature is computed by the volume part of the spatial metric, meaning that it does not contribute to the dispersion relation of GWs.
Indeed, expanding around FLRW background, we can prove that the graviton has the same dispersion relation as in GR. In fact, the tensor perturbation around FLRW background are obtained considering
\begin{equation}
N=1\,,\quad N^i=0\,,\quad \gamma_{ij}=a^2\left(\delta_{ij}+h_{ij}\right)\,,
\end{equation}
where $h_{ij}$ satisfies the transverse and traceless condition $\delta^{ij}\partial_i h_{jk}=0=\delta^{ij}h_{ij}$. Only the first two terms of Eq. \eqref{L_VCDM} (or \eqref{L_cus}) contribute to the dispersion relation, and these are the same as those that appear in the GR Lagrangian. The dispersion relation of GWs in the cuscuton/VCDM is the same as that in GR.

This concludes that the free propagation of gravitons in the cuscuton/VCDM is the same as that in GR. However, gravitational theories are non-linear theories and the gravitons are scattered via self-interactions. Our interest is whether the non-linear self-interactions are the same or not. To study the non-linear dynamics, we shall move to studying the Bianchi I universe.

\subsection{Bianchi I universe}
\label{sec_Bianchi_II}
Let us now consider the axisymmetric Bianchi I universe with 
\begin{align}
N=N(t)\,, \quad \lambda=\lambda(t)\,, \quad \phi=\phi(t)\,,
\end{align}
and
\begin{align}
\gamma_{ij}={\rm diag}\left[ a^2(t) e^{4\beta(t)}, a^2(t) e^{-2\beta(t)}, a^2(t)e^{-2\beta(t)} \right]
\,.
\end{align}
We then obtain
\begin{align}
K_{ij}K^{ij}-\frac{1}{3}K^2=6\frac{\dot{\beta}^2}{N^2}
\,, \quad 
K=3H
\,, \quad 
R=0\,,
\end{align}
where $H=\dot{a}/aN$. The minisuperspace action of the cuscuton/VCDM is 
\begin{align}
S_{\rm cus/VCDM}=\int dt Na^3\left[ 3\left(\frac{\dot{\beta}}{N} \right)^2-3H^2 -\mathscr{V}(3H) \right]
.
\end{align}

We emphasise that the effective gravitational constant for GWs in the cuscuton/VCDM is the same as GR because the coefficient in front of $R+K^{ij}K_{ij}-\frac{1}{3}K^2$ is $1/2$ (multiplied by the spacetime volume) in the unit with $M_{\rm Pl}=1$. This concludes that the equation of motion of the shear $\beta$ takes exactly the same form as that in GR:
\begin{align}
\frac{d}{dt}\left( a^3 \frac{\dot{\beta}}{N} \right)=0
\,.
\end{align}
The shear decreases as $a^{-3}$,
\begin{align}
\frac{\dot{\beta}}{N} \propto a^{-3}
\,.
\end{align}
Note that GWs are well-defined only in a high frequency/momentum limit compared with the background curvature scale. The combination $R+K^{ij}K_{ij}-\frac{1}{3}K^2$ already concludes that the propagation of GWs is the same as that in GR as we mentioned. Here, we showed that the dynamics of the low momentum part of ``GWs'', the shear, in the cuscuton/VCDM is also the same as in GR.

The cuscuton/VCDM only modifies the Friedmann equation as
\begin{align}
3H^2+3H\mathscr{V}'-\mathscr{V}=3\left(\frac{\dot{\beta}}{N} \right)^2
\,.
\end{align}
As a result, the expansion law of the universe is modified; in other word, the backreaction from the ``GWs'' to the background spacetime is modified in the cuscuton/VCDM. Intuitively, this modification would be caused by the non-local force carried by the cuscuton field, or the instantaneous/shadowy mode in more general terms~\cite{DeFelice:2018ewo}. If one interprets
\begin{align}
\rho_{\phi}=\mathscr{V}-3H\mathscr{V}'=V-\phi V_{\phi} +\frac{3}{2}V_{\phi}^2
\end{align}
as the energy density of the dark energy component, the cuscuton/VCDM theory realises the time-dependent dark energy without introducing a new dynamical degree of freedom~\cite{DeFelice:2020}.

Therefore, the modification of the cuscuton/VCDM theory only appears in the Friedmann equation while the equation of motion of the shear is unchanged. In the next section, we will show that this kind of modification cannot be achieved by the canonical transformation of GR, concluding that the cuscuton/VCDM theory is of type-II.

\section{Type-I MMG}\label{sec:type-I}

\subsection{Local canonical transformation}
We start with the Hamiltonian formalism of GR. The gravitational Hamiltonian of GR is given by
\begin{align}
H_{\rm tot}=\int d^3x (\Nold \mathcal{H}_0[\gamold,\piold] + \tilde{N}^i \mathcal{H}_i[\gamold,\piold] + \lambda \piold_N ) \,,
\label{H_GR}
\end{align}
with 
\begin{align}
\mathcal{H}_0 &:=\frac{2}{\sqrt{\gamold}}\left( \gamold_{i k} \gamold_{j l} -\frac{1}{2}\gamold_{ij} \gamold_{kl} \right) \piold^{ij}\piold^{kl}
-\frac{\sqrt{\gamold}}{2} R(\gamold)\,, \\
\mathcal{H}_i &:= -2\sqrt{\gamold} \gamold_{ij} \tilde{D}_k \left( \frac{\piold^{jk}}{\sqrt{\gamold}} \right) \,, 
\end{align}
where $\Nold,\tilde{N}^i, \gamold_{ij}$ are the lapse, the shift, and the spatial metric while $\piold_N$ and $\piold^{ij}$ are the canonical momenta for the lapse and the spatial metric, respectively.
Since we shall consider canonical transformations that explicitly preserve the spatial diffeomorphism invariance, the shift vector can be regarded as the Lagrange multiplier instead of the phase space variable. For this reason, we have not considered components of the shift vector as phase space variables and have not introduced the corresponding conjugate momenta. 
In the present paper, all variables with a tilde are the variables before a canonical transformation, namely the variables in the Einstein frame. On the other hand, we will use variables without a tilde to denote the variables in the Jordan frame which matter fields minimally couple with.

By definition, the Hamiltonian of any type-I MMG theories is obtained from \eqref{H_GR} via an invertible transformation of the variables. In the phase space, the most general transformation of the variables is a canonical transformation, which is specified by a generating functional, 
\begin{align}
F=-\int d^3x \sqrt{\gamma} \mathcal{F}
\,. \label{gen_func}
\end{align}
Here, the generating function is supposed to be a function of old canonical momenta $(\piold_N,\piold^{ij})$ and new variables $(N,\gamma_{ij})$. As already stated, components of the shift vector have been regarded as Lagrange multipliers instead of phase space variables. 
We now impose the following two conditions to {\it define} type-I MMG theories in a more proper way:
\begin{enumerate}
\item the generating function does not depend on the time $t$ explicitly and
\item the generating function $\mathcal{F}$ is a scalar under the spatial diffeomorphism.
\end{enumerate}
The first condition is imposed in order that the transformation is local in the time domain whereas the second condition is imposed to preserve the spatial diffeomorphism invariance explicitly. Note that the canonical transformation itself would not violate the symmetry of the theory and thus the spatial diffeomorphism invariance is preserved in a non-trivial way even if we consider a non-scalar generating function. However, in this case, the new metric $\gamma_{ij}$ is no longer a spatial tensor and the spatial diffeomorphism invariance should be broken when a matter field is introduced and minimally coupled to the metric after the canonical transformation (see~\cite{Carballo-Rubio:2018,Aoki:2018zcv}). We thus demand that the spatial diffeomorphism invariance is preserved {\it explicitly}, that is, all new variables are spatial tensors (or spatial tensor densities). We also note that our transformation may include the spatial (covariant) derivatives of the canonical variables. In such a case, the transformation is generically non-local in the spatial domain but the transformation may be invertible under an appropriate boundary condition.

The most general type-I MMG theory is obtained by considering a generating function of the form
\begin{align}
\mathcal{F}=\mathcal{F}(\piold_N/\sqrt{\gamma},\piold^{ij}/\sqrt{\gamma},N,\gamma_{ij},R^i{}_{jkl}(\gamma), D_i)
\end{align}
where $R^i{}_{jkl}(\gamma)$ and $D_i$ are the Riemann tensor and the spatial covariant derivative compatible with the metric $\gamma_{ij}$, respectively. We put the $1/\sqrt{\gamma}$ factors in the canonical momenta because the canonical momenta are not tensors while $\piold_N/\sqrt{\gamma},\piold^{ij}/\sqrt{\gamma}$ are tensors.

From the first derivatives of the generating function, we obtain relations between the new canonical variables and the old canonical variables. The transformed Hamiltonian in terms of the new variables is obtained by substituting the solutions, $\piold_N({\rm new}), \piold^{ij}({\rm new}), \Nold({\rm new}),\gamold_{ij}({\rm new})$ where $({\rm new})$ means that the quantities are functions of the new canonical variables. Then, as far as we keep the last term in \eqref{H_GR}, we can freely change/introduce/eliminate the $\piold_N({\rm new})$ dependency in the remaining part of the Hamiltonian by redefining the Lagrange multiplier $\lambda$. Therefore, it is sufficient to expand the generating function up to linear order in $\piold_N$,
\begin{align}
\mathcal{F}=f(\piold^{ij}/\sqrt{\gamma},N,\gamma_{ij},R^i{}_{jkl}(\gamma), D_i) + g (\piold^{ij}/\sqrt{\gamma},N,\gamma_{ij},R^i{}_{jkl}(\gamma), D_i) \piold_N/\sqrt{\gamma}
\,,
\label{general_gen}
\end{align}
because any effects from terms non-linear in $\piold_N$ can be eliminated by redefining the Lagrange multiplier $\lambda$.

\subsection{Relative propagation speed of GWs}
As we mentioned, MMG theories can be classified into type-I and type-II~\cite{Aoki:2018}. After adding a matter field, say photon, each MMG type can be further classified into two subclasses: (a) the propagation of GWs is the same as electromagnetic waves in a high energy limit; and (b) the relative propagations are different, where the high energy limit means that frequencies of the waves are sufficiently higher than the background curvature scale. As a result, we have four subclasses of MMG theories which we shall call type-Ia, Ib, IIa, and IIb, respectively. Clearly, theories studied in Sec.~V of ref.~\cite{Aoki:2018} are examples of the type-Ia MMG theories and the cuscuton/VCDM theory is a candidate for the type-IIa MMG theories.

\begin{figure}[t] 
\centering
 \includegraphics[width=0.75\linewidth]{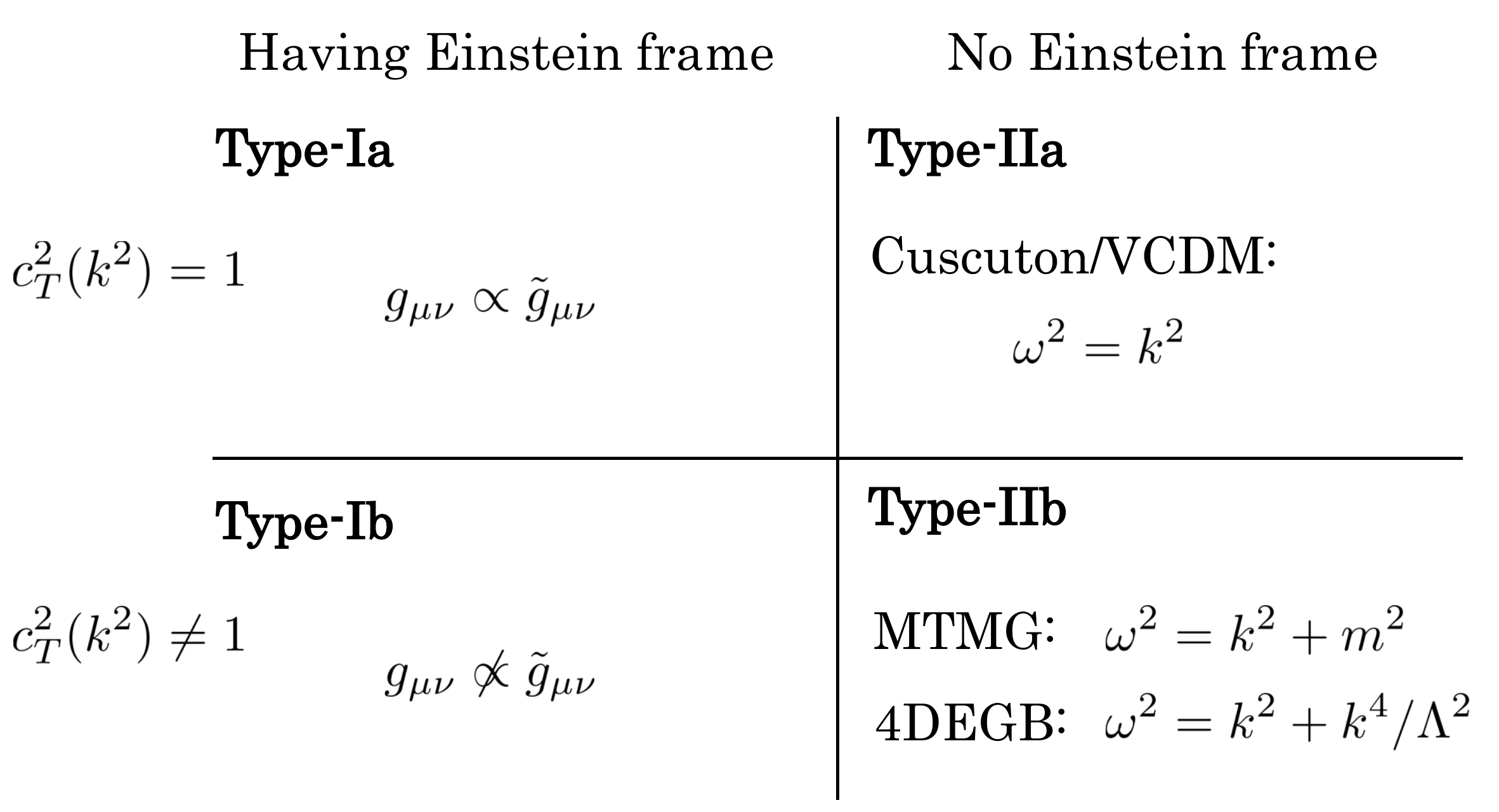}	
 \caption{Classifications of MMG. As for type-I MMG we show how the Einstein frame metric and the Jordan frame metric are related while the dispersion relations of GWs are shown for each example type-II MMG theory where MTMG stands for the minimal theory of massive gravity~\cite{DeFelice:2015hla} and 4DEGB is the consistent $D\to 4$ Einstein-Gauss-Bonnet gravity~\cite{Aoki:2020lig}.}
 \label{fig_class}
\end{figure}

The canonical transformation does not change the form of the dispersion relation (see Appendix \ref{app_can_scalar}, where we investigate a canonical transformation generated by a generating functional that in general involves spatial derivatives), meaning that the MMG theories with the dispersion relations like $\omega^2=k^2+m^2$~\cite{DeFelice:2015hla} or $\omega^2=k^2+k^4/\Lambda^2$~\cite{Aoki:2020lig} are classified into the type-IIb MMG theories. One may then wonder why we need to define the type-Ib theories if the dispersion relation is invariant under the canonical transformations. As is well-known in the context of scalar-tensor theories, the light cone is invariant under conformal transformations whereas disformal transformations change the light cone, modifying the relative propagations of GWs and the matter fields. The type-Ia MMG theories are related to GR via a conformal type canonical transformation and the type-Ib MMG theories are other type-I theories, that is, the Einstein frame metric and the Jordan frame metric are related in a disformal way. The summary of the classification is shown in Fig.~\ref{fig_class}.

Let us elaborate the (non-)invariance of the propagation of GWs under the canonical transformation relative to the propagation of matter. The canonical transformations cause the change of the metric
\begin{align}
\tilde{g}_{\mu\nu }(\tilde{N},\tilde{N}^i, \gamold_{ij}) \to g_{\mu\nu}(N,N^i, \gamma_{ij})\,, \quad
 \tilde{N}^i = N^i\,,
\end{align}
where $N$ and $\gamma_{ij}$ are functions of $\tilde{N}$, $\gamold_{ij}$ and their derivatives. The matter fields are supposed to be coupled with the Jordan frame metric $g_{\mu\nu}$ in MMG after adding a gauge-fixing condition~\cite{Aoki:2018zcv}. For instance, we consider a test massless scalar field $\psi$ as a matter field. The equation of motion is
\begin{align}
g^{\mu\nu}\nabla_{\mu}\nabla_{\nu} \psi=0
\,,
\end{align}
which is approximated by
\begin{align}
g^{\mu\nu}\partial_{\mu}\partial_{\nu} \psi =0
\,,
\end{align}
in the high energy limit. On the other hand, by definition, the equation of motion of GWs in any type-I MMG is the same as that in GR in the Einstein frame; in the high energy limit,
\begin{align}
\tilde{g}^{\mu\nu}\partial_{\mu}\partial_{\nu}\tilde{h}^{TT}_{ij}=0
\,,
\label{pro_Einstein}
\end{align}
where $\tilde{g}^{\mu\nu}$ is understood as the background (low frequency) part of the Einstein frame metric and $\tilde{h}^{TT}_{ij}$ is the perturbation describing GWs. Although we consider the equation of motion for the perturbations of the Einstein frame metric in \eqref{pro_Einstein} rather than the Jordan frame one, it does not matter for the present discussion (see Appendix \ref{app_can_scalar}). The point is that the field $\psi$ propagates on the Jordan frame metric $g^{\mu\nu}$ while the GW propagates on the Einstein frame metric $\tilde{g}^{\mu\nu}$. The change of the background metric $\tilde{g}_{\mu\nu } \to g_{\mu\nu} $ deviates the relative propagation of GW from the propagation of the matter field. For example, around the FLRW background with the lapses $\tilde{N},N$ and the scale factors $\tilde{a},a$, the dispersion relation of the matter field is
\begin{align}
\frac{\omega^2}{N^2}-\frac{k^2}{a^2} =0
\,,
\end{align}
while that of GW is
\begin{align}
\frac{\omega^2}{\tilde{N}^2}-\frac{k^2}{\tilde{a}^2}=\frac{N^2}{\tilde{N}^2}\left( \frac{\omega^2}{N^2}- c_T^2 \frac{k^2}{a^2}  \right)=0
\,,
\end{align}
where
\begin{align}
c_T^2=\frac{a^2/N^2}{\tilde{a}^2/\tilde{N}^2}
\,.
\end{align}
The sound speed of GWs is different from unity, in general.

In the type-Ia MMG theories the Jordan frame metric is conformally related to that in the Einstein frame, $g_{\mu\nu}\propto \tilde{g}_{\mu\nu}$. Then, the equation \eqref{pro_Einstein} is reduced to
\begin{align}
g^{\mu\nu}\partial_{\mu}\partial_{\nu}\tilde{h}^{TT}_{ij}=0
\end{align}
in the type-Ia MMG. The propagation of GW is the same as that of the matter fields, which are coupled with the Jordan frame metric, in the high energy limit.

\subsection{Type-Ia MMG}
Hereinafter, we focus on MMG theories which have the same dispersion relation as photon in the Jordan frame, namely type-Ia MMG theories. Such a subclass of the type-I MMG theories is generated by a conformal canonical transformation from GR.

Without loss of generality, we can suppose that the generating function does not involve spatial derivatives because we will focus on the Bianchi I universe, which is spatially homogeneous. From the spatial covariance, the functions $f$ and $g$ in \eqref{general_gen} are given by
\begin{align}
f=f(\Phi^i{}_j,N)\,, \quad g=g(\Phi^i{}_j,N)
\end{align}
where 
\begin{align}
\Phi^i{}_j=\tilde{\pi}^{ik}\gamma_{kj}/\sqrt{\gamma}
\,.
\end{align}
Adopting the notations
\begin{align}
f_{i}{}^j=\frac{\partial f}{\partial \Phi^i{}_j} \,, \quad f_{ij}=f_{i} {}^k \gamma_{kj}
\,, \quad f_{N}=\frac{\partial f}{\partial N}
\,, \quad g_{N}=\frac{\partial g}{\partial N}
\,,
\end{align}
the old metric and the old lapse function are given by
\begin{align}
\tilde{\gamma}_{ij}&=f_{ij}
\,, 
\label{old_met} \\
\tilde{N}&=g
\,,
\end{align}
where we have neglected the term coming from $g$ in \eqref{old_met} because this term is proportional to $\piold_N$ and thus its contributions to the Hamiltonian can be eliminated by redefining the Lagrange multiplier $\lambda$.

We first demand that the spatial metric is transformed in a conformal way, i.e. $f_{ij}\propto \gamma_{ij}$. This is a necessary condition for the two spacetime metrics $\tilde{g}_{\mu\nu}$ and $g_{\mu\nu}$ to be conformally related to each other. The spatial covariance and Cayley-Hamilton theorem imply that $f$ is a function of the scalar quantities $\Phi^i{}_i, \Phi^i{}_j \Phi^j{}_i, \Phi^i{}_j \Phi^j{}_k \Phi^k{}_i$ and $N$. If $f$ depends on either $\Phi^i{}_j \Phi^j{}_i$ or $\Phi^i{}_j \Phi^j{}_k \Phi^k{}_i$, the derivative $f^i{}_j$ is no longer proportional to $\delta^i{}_j$ and then the transformation is not conformal. Therefore, the necessary condition for the canonical transformation to be conformal is 
\begin{align}
f=f(\Phi,N)
\end{align}
where $\Phi=\Phi^i{}_i$. The old spatial metric and the new spatial metric are related by
\begin{align}
\gamold_{ij}=f_{\Phi}\gamma_{ij}\,, \quad f_{\Phi}=\frac{\partial f}{\partial \Phi}
\,.
\end{align}
Next, we demand that the spacetime metrics, 
\begin{align}
\tilde{g}_{\mu\nu}dx^{\mu}dx^{\nu}&=-\tilde{N}^2dt^2 + \gamold_{ij}(dx^i+\tilde{N}^idt)(dx^j+\tilde{N}^jdt)
\,,\nonumber\\
g_{\mu\nu}dx^{\mu}dx^{\nu}&=-N^2dt^2 + \gamma_{ij}(dx^i+N^idt)(dx^j+N^jdt)
\,,
\end{align}
are also transformed conformally to each other. By this requirement, the shift vectors $N^i$ and $\tilde{N}^i$ have to be identical, $N^i=\tilde{N}^i$, and the function $g$ is determined to be
\begin{align}
g=f_{\Phi}^{1/2}N
\,.
\end{align}
As a result, the generating functional for the conformal canonical transformation is given by
\begin{align}
F=\int d^3x \left[ \sqrt{\gamma} f(\Phi,N)+ f_{\Phi}^{1/2}(\Phi,N) N \piold_N \right]
.
\end{align}
With this form for the generating functional we have
\begin{equation}
\pi_N=\sqrt{\gamma}f_N+\frac{1}{2}f_{N\Phi}f_{\Phi}^{-1/2}N\tilde{\pi}_{\tilde{N}}+f_{\Phi}^{1/2}\tilde{\pi}_{\tilde{N}}
\end{equation}
and
\begin{equation}
\pi^{ij}=f_\Phi\tilde{\pi}^{ij}+\frac{1}{2}\sqrt{\gamma}\gamma^{ij}\left(f-\Phi f_{\Phi}\right)
.
\end{equation}

The transformed Hamiltonian is explicitly computed as follows. We regard the variable $\Phi$ and its conjugate $p$ as independent variables in the phase space being subject to the constraints
\begin{align}
p\approx 0 \,, \quad
\pi-\frac{\sqrt{\gamma}}{2}(3f-f_{\Phi}\Phi) \approx 0
\,.
\end{align}
The old variables and the new variables are related via
\begin{align}
\gamold_{ij}&=f_{\Phi} \gamma_{ij} \,, \\
\Nold&=f_{\Phi}^{1/2}N \,, \\
\piold^{ij}&=f_{\Phi}^{-1}\left[ \pi^{ij} -\frac{1}{2}\sqrt{\gamma}\gamma^{ij}(f-\Phi f_{\Phi}) \right]
\,, \\
\piold_N&=f_{\Phi}^{-1/2}\left(1+\frac{f_{N\Phi}}{2f_{\Phi}}\right)^{-1}(\pi_N-\sqrt{\gamma} f_N)
\,.
\end{align}
After redefining the Lagrange multipliers, the transformed Hamiltonian is given by
\begin{align}
H_{\rm tot}=\int d^3x \Biggl[ &N \mathcal{H}_0[\gamma,\pi,\Phi,N] + N^i \mathcal{H}_i[\gamma,\pi] + \lambda (\piold_N-\sqrt{\gamma}f_N)  
\nn
&+\lambda_p p+\lambda_{\Phi}\left(\pi-\frac{\sqrt{\gamma}}{2}(3f-f_{\Phi}\Phi) \right) \Biggl] \,,
\label{H_typeIa}
\end{align}
where $\lambda_p$ and $\lambda_{\Phi}$ are Lagrange multipliers implementing the constraints and
\begin{align}
\mathcal{H}_0 =&\frac{2}{f_{\Phi} \sqrt{\gamma}}\left( \gamma_{i k} \gamma_{j l} -\frac{1}{2}\gamma_{ij} \gamma_{kl} \right) \pi^{ij}\pi^{kl}
+\frac{\sqrt{\gamma}}{4f_{\Phi}}(f-\Phi f_{\Phi})(3f+\Phi f_{\Phi})
\nn
&-\frac{f_{\Phi}\sqrt{\gamma}}{2} \left( R(\gamma) -2D^2 \ln f_{\Phi}-\frac{1}{2}(D\ln f_{\Phi})^2 \right)
, \\
\mathcal{H}_i =& -2\sqrt{\gamma} \gamma_{ij} D_k \left( \frac{\pi^{jk}}{\sqrt{\gamma}} \right) 
.
\end{align}
The corresponding Lagrangian is obtained by taking the Legendre transformation.\footnote{One should add a gauge-fixing condition, say $\lambda_{\rm gf}^i \partial_i \Phi$ before taking the Legendre transformation in order to introduce matter fields~\cite{Aoki:2018zcv} which is, nonetheless, irrelevant to the present purpose because we will only consider a vacuum solution.} After integrating out $\lambda_{\Phi}$, the Lagrangian is
\begin{align}
\mathcal{L}_{\text{type-Ia}}=&\frac{f_{\Phi}}{2}N\sqrt{\gamma}\left[ 
K^{ij}K_{ij}-\frac{1}{3}K^2+   R(\gamma) -2D^2 \ln f_{\Phi}-\frac{1}{2}(D\ln f_{\Phi})^2 \right]
\nn
&+N\sqrt{\gamma}\left[ K\left(f-\frac{1}{3}\Phi f_{\Phi} \right) +\frac{1}{3}\Phi^2 f_{\Phi} +f_N \frac{d}{dt} \ln N\right]
\,.
\end{align}

\subsection{Bianchi I universe}
Let us now restrict our attention to the Bianchi I universe. Due to the spatial homogeneity, we can discard terms containing spatial derivatives, $D^2 \ln f_{\Phi}$ and $D\ln f_{\Phi}$.
Also, this restriction justifies our assumption that the generating functional of the canonical transformation does not involve spatial derivatives. 

The point is that the effective gravitational constant for GWs in the type-Ia MMG is in general different from that in the cuscuton/VCDM. The minisuperspace action of the type-Ia MMG theory is
\begin{align}
S_{\text{type-Ia}}= \int dt Na^3  \left[ 3f_{\Phi} \left(\frac{\dot{\beta}}{N} \right)^2
+3H\left(f-\frac{1}{3}\Phi f_{\Phi} \right)+\frac{1}{3}\Phi^2 f_{\Phi}+f_N \frac{d}{dt}\ln N
\right]
.
\label{mini_S_Ia}
\end{align}
The equation of motion of the shear is
\begin{align}
\frac{d}{dt} \left( f_{\Phi}a^3 \frac{\dot{\beta}}{N} \right)=0
\,,
\label{eq_beta_Ia}
\end{align}
leading to the decaying law $\dot{\beta}/N \propto f_{\Phi}^{-1}a^{-3}$. To reproduce the decaying behaviour of the shear in the cuscuton/VCDM, one should choose $f_{\Phi}={\rm constant}$ which is the constant rescaling of the metric and does not modify the Friedmann equation. This concludes that the cuscuton/VCDM is not a type-Ia MMG theory; that is, the cuscuton/VCDM is a type-IIa MMG theory.

To confirm the consistency of our calculations, let us investigate the Friedmann equation in the type-Ia MMG theory. Variations with respect to $\Phi$ and $a$ provide
\begin{align}
H&=-\frac{\Phi (2f_{\Phi}+\Phi f_{\Phi \Phi})+9f_{\Phi \Phi} (\dot{\beta}/N)^2+3 f_{\Phi N} \dot{N}/N }{3(2f_{\Phi}-\Phi f_{\Phi\Phi})}
\,,
\label{eq_H}\\
\frac{\dot{\Phi}}{N}&= \frac{\Phi^2 f_{\Phi}+9 f_{\Phi}(\dot{\beta}/N)^2+f_{\Phi N} \Phi \dot{N}/N}{2f_{\Phi}-\Phi f_{\Phi\Phi}}
\,.
\label{eq_dPhi}
\end{align}
By the use of these expressions, variation with respect to $N$ yields
\begin{align}
\Phi^2=9\left( \frac{\dot{\beta}}{N} \right)^2
\,. \label{Fri_Ia}
\end{align}
Taking derivative of \eqref{Fri_Ia} with respect to $t$ and using \eqref{eq_beta_Ia}, \eqref{eq_H}, and \eqref{Fri_Ia}, one can obtain \eqref{eq_dPhi}. Hence, the independent equations of the system are \eqref{eq_beta_Ia}, \eqref{eq_H}, and \eqref{Fri_Ia}. The dynamics of $N$ is undermined from the equations, reflecting the fact that \eqref{mini_S_Ia} is invariant under the time parametrization, $t\to t'=t'(t)$. One can choose the time parameter so that $N=1$. Then, $\Phi$ can be given as a function of $H$ and $\dot{\beta}$ by solving the algebraic equation \eqref{eq_H}. The Friedmann equation is obtained by substituting $\Phi=\Phi(H,\dot{\beta})$ into \eqref{Fri_Ia}. Note that one can show the relation
\begin{align}
3\tilde{H}^2-3\left(\frac{\dot{\beta}}{\tilde{N}} \right)^2 \propto \Phi^2-9\left( \frac{\dot{\beta}}{N} \right)^2
\end{align}
by means of \eqref{eq_H} and \eqref{eq_dPhi}, where
\begin{align}
\tilde{H}=\frac{\dot{\tilde{a}}}{\tilde{a}\tilde{N}}
\,, \quad
\tilde{a}=f_{\Phi}^{1/2}a\,, \quad
\tilde{N}=f_{\Phi}^{1/2}N
\,,
\end{align}
are the Einstein frame variables. This confirms that the present system is obtained by a change of variables from GR.
If $f_{\Phi}={\rm constant}$, the equation \eqref{eq_H} is reduced to
\begin{align}
H=-\frac{1}{3}\Phi
\end{align}
and then the standard Friedmann equation,
\begin{align}
3H^2=3\left( \frac{\dot{\beta}}{N} \right)^2
\,,
\end{align}
is obtained.

As we have seen in \S.~\ref{sec_Bianchi_II}, the cuscuton/VCDM theory only modifies the Friedmann equation while preserving the equation of motion of the shear. On the other hand, the type-Ia MMG theory changes both of the equations. As a result, the dynamics of the cuscuton/VCDM is not reproduced by type-Ia MMG, meaning that the cuscuton/VCDM cannot be obtained by a field transformation from GR. The cuscuton/VCDM is an explicit example for a gravitational theory in which the number of dynamical degrees of freedom is the same as GR but which is inequivalent to GR even in vacuum.

\section{Summary and Discussions}
In this paper we have analyzed a specific theory of minimally modified gravity called VCDM. We have proved that this theory is equivalent to the cuscuton theory and that both are type-II theories. In fact, even in the absence of matter, it is not possible to recover the dynamic of the theory via a field redefinition of the GR action.

In order to prove it, it was useful to introduce a refined classification of MMG theories by means of type-a and type-b theories. 
In type-a theories gravitational waves have the usual dispersion relation, whereas in type-b theories the dispersion relation is modified. To our knowledge, the proof that cuscuton and VCDM do not have an Einstein frame constitutes the first proof of the existence of type-IIa theories. (On the other hand, examples of type-IIb theories have been known~\cite{DeFelice:2015hla,Aoki:2020lig,Yao:2020tur}.) This is very relevant as modified dispersion relations are constrained by observations~\cite{Abbott_2017}~\footnote{For example, the graviton mass in MTMG~\cite{DeFelice:2015hla} and the coefficient of the $k^4$ term in 4DEGB~\cite{Aoki:2020lig} and its generalization~\cite{Yao:2020tur} are mildly constrained. On the other hand, the deviation of the coefficient of the $k^2$ term from the squared speed of light is strongly constrained.}. As a byproduct of our analysis, we have obtained the most generic type-Ia theory by considering conformal canonical transformation of the GR action, as far as spatially homogeneous solutions are concerned. Furthermore, dropping the requirement of having a conformal transformation, the analysis can be easily extended to obtain more general type-I theories. 

\begin{acknowledgments}
\noindent
The work of K.A. was supported in part by Grants-in-Aid from the Scientific Research Fund of the Japan Society for the Promotion of Science, No. 19J00895 and No. 20K14468.\\
The work of F.D.F. was supported by Japan Society for the Promotion of Science Grants-in-Aid for Scientific Research No.~17H06359.\\
The work of S.M.\ was supported in part by Japan Society for the
Promotion of Science Grants-in-Aid for Scientific Research
No.~17H02890, No.~17H06359, and by World Premier International
Research Center Initiative, MEXT, Japan.
\end{acknowledgments}
\appendix

\section{Canonical transformation of a free scalar}
\label{app_can_scalar}

For simplicity we consider a free massless scalar field $\phiold$ in flat spacetime described by the Hamiltonian in the momentum space,
\begin{align}
H_{\rm scalar}=\int d^3k \left[ \frac{1}{2}\piold^2+\frac{1}{2}k^2 \phiold^2 \right]\,.
\end{align}
The most general linear canonical transformation is specified by a generating functional of the form
\begin{align}
F=-\int d^3k \left(\frac{1}{2}f_{\phi\phi}\phi\phi+f_{\phi\pi}\phi \piold + \frac{1}{2}f_{\pi\pi}\piold \piold \right)\,,
\end{align}
where the coefficients $f_{\phi\phi}, f_{\phi\pi}, f_{\pi\pi}$ are functions of $k^2$ because we are allowing for a generating functional that contains spatial derivatives. For each Fourier component, the canonical transformation is given by
\begin{align}
\phiold=f_{\phi \pi} \phi+ f_{\pi \pi} \piold\,, \quad
\pi=f_{\phi\phi}\phi +f_{\phi \pi} \piold
\,,
\label{can_free_s}
\end{align}
which can be solved with respect to the old variables as 
\begin{align}
\piold=\frac{1}{f_{\phi \pi}} (\pi- f_{\phi\phi} \phi)
\,, \quad
\phiold=( f_{\phi\pi}^2- f_{\pi\pi} f_{\phi\phi} ) \frac{\phi}{f_{\phi \pi}}+\frac{f_{\pi \pi}}{f_{\phi \pi}} \pi
\,,
\end{align}
and the resultant Hamiltonian is no longer proportional to $\pi^2+k^2 \phi^2$. However, one should discuss the equation of motion to see the dispersion relation of the new variable $\phi$. The Hamilton's equations for the new canonical variables are
\begin{align}
\dot{\pi}&=\left[ f_{\phi\phi}-f_{\pi\pi}(f_{\phi\pi}^2- f_{\pi\pi} f_{\phi\phi})k^2 \right] \frac{\pi}{f_{\phi\pi}^2} -\left[ f_{\phi\phi}^2+(f_{\phi\pi}^2- f_{\pi\pi} f_{\phi\phi} )^2k^2 \right] \frac{\phi}{f_{\phi \pi}^2}
\,,\\
\dot{\phi}&=(1+f_{\pi\pi}^2 k^2) \frac{\pi}{f_{\phi \pi}^2} - \left[ f_{\phi\phi}-f_{\pi\pi} (f_{\phi\pi}^2- f_{\pi\pi} f_{\phi\phi})k^2\right] \frac{\phi}{f_{\phi\pi}^2}
\,,
\end{align}
from which we find
\begin{align}
\ddot{\phi}+k^2 \phi=0
\,. \label{dis_s}
\end{align}
The new variable $\phi$ obeys the same dispersion relation as the old variable.

The invariance of the dispersion relation is almost trivial from the relation \eqref{can_free_s}. By the use of the Hamilton's equations for the old variables, the first equation of \eqref{can_free_s} becomes
\begin{align}
\phi=\frac{1}{f_{\phi \pi}}(
\phiold-f_{\pi\pi}\dot{\phiold})
\,.
\end{align}
One can easily obtain \eqref{dis_s} by using the old equation of motion $\ddot{\phiold}+k^2\phiold=0$.

While in the above we have studied a scalar field in the flat background, the same conclusion about the dispersion relation holds also in curved backgrounds. This is because a dispersion relation is well-defined only for modes with momenta sufficiently higher than the background curvature scale.

\newpage
\bibliographystyle{ieeetr}
\bibliography{refs}

\begin{thebibliography}{10}

\bibitem{Lin:2017}
C.~Lin and S.~Mukohyama, ``{A Class of Minimally Modified Gravity Theories},''
  {\em JCAP}, vol.~10, p.~033, 2017.

\bibitem{Lin:2019}
C.~Lin and Z.~Lalak, ``{Novel matter coupling in Einstein gravity},'' 11 2019.

\bibitem{Carballo-Rubio:2018}
R.~Carballo-Rubio, F.~Di~Filippo, and S.~Liberati, ``{Minimally modified
  theories of gravity: a playground for testing the uniqueness of general
  relativity},'' {\em JCAP}, vol.~06, p.~026, 2018.
\newblock [Erratum: JCAP 11, E02 (2018)].

\bibitem{Aoki:2018}
K.~Aoki, A.~De~Felice, C.~Lin, S.~Mukohyama, and M.~Oliosi, ``{Phenomenology in
  type-I minimally modified gravity},'' {\em JCAP}, vol.~01, p.~017, 2019.

\bibitem{DeFelice:2020}
A.~De~Felice, A.~Doll, and S.~Mukohyama, ``{A theory of type-II minimally
  modified gravity},'' {\em JCAP}, vol.~09, p.~034, 2020.

\bibitem{DeFelice:2020cpt}
A.~De~Felice, S.~Mukohyama, and M.~C. Pookkillath, ``{Addressing $H_0$ tension
  by means of VCDM},'' {\em Phys. Lett. B}, vol.~816, p.~136201, 2021.

\bibitem{Afshordi:2006ad}
N.~Afshordi, D.~J.~H. Chung, and G.~Geshnizjani, ``{Cuscuton: A Causal Field
  Theory with an Infinite Speed of Sound},'' {\em Phys. Rev. D}, vol.~75,
  p.~083513, 2007.

\bibitem{Lovelockth}
D.~Lovelock, ``The einstein tensor and its generalizations,'' {\em Journal of
  Mathematical Physics}, vol.~12, no.~3, pp.~498--501, 1971.

\bibitem{Gupta}
S.~N. Gupta, ``Einstein's and other theories of gravitation,'' {\em Rev. Mod.
  Phys.}, vol.~29, pp.~334--336, Jul 1957.

\bibitem{Deser1969}
S.~Deser, ``{Selfinteraction and gauge invariance},'' {\em Gen. Rel. Grav.},
  vol.~1, pp.~9--18, 1970.

\bibitem{Padmanabhan2004}
T.~Padmanabhan, ``{From gravitons to gravity: Myths and reality},'' {\em Int.
  J. Mod. Phys.}, vol.~D17, pp.~367--398, 2008.

\bibitem{Carballo-Rubio:2018b}
R.~Carballo-Rubio, F.~Di~Filippo, and N.~Moynihan, ``{Taming higher-derivative
  interactions and bootstrapping gravity with soft theorems},'' {\em JCAP},
  vol.~10, p.~030, 2019.

\bibitem{Britto2005}
R.~Britto, F.~Cachazo, B.~Feng, and E.~Witten, ``Direct proof of the tree-level
  scattering amplitude recursion relation in yang-mills theory,'' {\em Phys.
  Rev. Lett.}, vol.~94, p.~181602, May 2005.

\bibitem{Benincasa2007}
P.~Benincasa, C.~Boucher-Veronneau, and F.~Cachazo, ``{Taming Tree Amplitudes
  In General Relativity},'' {\em JHEP}, vol.~11, p.~057, 2007.

\bibitem{ArkaniHamed2008}
N.~Arkani-Hamed and J.~Kaplan, ``{On Tree Amplitudes in Gauge Theory and
  Gravity},'' {\em JHEP}, vol.~04, p.~076, 2008.

\bibitem{Pajer:2020}
E.~Pajer, D.~Stefanyszyn, and J.~Supe\l{}, ``{The Boostless Bootstrap:
  Amplitudes without Lorentz boosts},'' {\em JHEP}, vol.~12, p.~198, 2020.

\bibitem{Benincasa:2013}
P.~Benincasa, ``{New structures in scattering amplitudes: a review},'' {\em
  Int. J. Mod. Phys. A}, vol.~29, no.~5, p.~1430005, 2014.

\bibitem{Conde:2014}
E.~Conde, ``{Physics from the S-matrix: Scattering Amplitudes without
  Lagrangians},'' {\em PoS}, vol.~Modave 2013, p.~005, 2014.

\bibitem{DeFelice:2020onz}
A.~De~Felice, A.~Doll, F.~Larrouturou, and S.~Mukohyama, ``{Black holes in a
  type-II minimally modified gravity},'' {\em JCAP}, vol.~03, p.~004, 2021.

\bibitem{DeFelice:2018ewo}
A.~De~Felice, D.~Langlois, S.~Mukohyama, K.~Noui, and A.~Wang, ``{Generalized
  instantaneous modes in higher-order scalar-tensor theories},'' {\em Phys.
  Rev. D}, vol.~98, no.~8, p.~084024, 2018.

\bibitem{Iyonaga:2018vnu}
A.~Iyonaga, K.~Takahashi, and T.~Kobayashi, ``{Extended Cuscuton:
  Formulation},'' {\em JCAP}, vol.~12, p.~002, 2018.

\bibitem{Pajer:2018egx}
E.~Pajer and D.~Stefanyszyn, ``{Symmetric Superfluids},'' {\em JHEP}, vol.~06,
  p.~008, 2019.

\bibitem{Grall:2019qof}
T.~Grall, S.~Jazayeri, and E.~Pajer, ``{Symmetric Scalars},'' {\em JCAP},
  vol.~05, p.~031, 2020.

\bibitem{DeFelice:2020prd}
A.~De~Felice and S.~Mukohyama, ``{Weakening gravity for dark matter in a
  type-II minimally modified gravity},'' 11 2020.

\bibitem{Aoki:2018zcv}
K.~Aoki, C.~Lin, and S.~Mukohyama, ``{Novel matter coupling in general
  relativity via canonical transformation},'' {\em Phys. Rev. D}, vol.~98,
  no.~4, p.~044022, 2018.

\bibitem{DeFelice:2015hla}
A.~De~Felice and S.~Mukohyama, ``{Minimal theory of massive gravity},'' {\em
  Phys. Lett. B}, vol.~752, pp.~302--305, 2016.

\bibitem{Aoki:2020lig}
K.~Aoki, M.~A. Gorji, and S.~Mukohyama, ``{A consistent theory of $D \to 4$
  Einstein-Gauss-Bonnet gravity},'' {\em Phys. Lett. B}, vol.~810, p.~135843,
  2020.

\bibitem{Yao:2020tur}
Z.-B. Yao, M.~Oliosi, X.~Gao, and S.~Mukohyama, ``{Minimally modified gravity
  with an auxiliary constraint: A Hamiltonian construction},'' {\em Phys. Rev.
  D}, vol.~103, no.~2, p.~024032, 2021.

\bibitem{Abbott_2017}
B.~P. Abbott~\textit{et al.}, ``Gravitational waves and gamma-rays from a
  binary neutron star merger: {GW}170817 and {GRB} 170817a,'' {\em The
  Astrophysical Journal}, vol.~848, p.~L13, oct 2017.

\end{thebibliography}

\end{document}